# Folding Catastrophes due to Viscosity in Multiferroic Domains: Implications for Room-temperature Multiferroic Switching

J. F. Scott*

*School of Chemistry and School of Physics and Astronomy*

*St. Andrews University, St. Andrews, KY16 9ST, U.K.*

jfs4@st-andrews.ac.uk

1. Introduction

Room-temperature studies of magnetic switching of ferroelectric domains and electric switching of magnetic domains were reported [1] in the single-phase but complicated perovskite oxide $Pb[Fe_{1/2}Ta_{1/2}]_{1-x}[Ti_{1-y}Zr_y]_xO_3$ by Evans et al. in 2013, a paper downloaded more than 10,000 times in two years. In addition to providing a potentially new kind of memory device, this work displayed some deep puzzles: (1) Although the magnetic and ferroelectric domains switched polarization/magnetization under applied fields, they failed to satisfy the Landau-Lifshitz-Kittel Law [2] for domain widths versus film thickness; (2) the domain walls were not rectilinear but instead were highly curved, resembling grapefruit sections; (3) the polarization switching with magnetic fields H fatigued quickly (a few cycles), whereas ferroelectric domain switching with E was indefinitely repetitive. These observations are now reviewed in terms of the theory of folding catastrophes in magnetic domains.[3,4] The present interpretation is that due to domain wall viscosity $\eta$ [5] there is an instability at a threshold in mean wall velocity v at which domain walls fold and slide over each other, causing the Kittel Law to fail and displaying lava-like high-viscosity folded patterns. For ferroelectric domains the earlier work of Dawber [5] shows quantitatively that motion is ballistic in a high-viscosity medium, with viscosity due to acoustic phonon scattering.

2. Data [1]

In addition to the curved surfaces and failure to satisfy the Kittel Law, the rounded "domains" measured at Queen's University Belfast [1] via TEM and PFM techniques on focused-ion-beam cut nano-crystals reveal a quantitative agreement with folding catastrophes in high-viscosity materials: Figures 1a,b show that their maximum widths are ca. 100 nm at a film thickness of 150 nm cut by focused-ion beam (FIB). Figure 2 (thicker) is slightly smoother with a more regular wavelength of folds. In theoretical studies [6] of the threshold for buckling compared with smooth folding, there is a characteristic ratio of the "dominant wavelength" w (corresponding to our widths) to the film thickness d, and this ratio, unlike the Landau-Lifshitz-Kittel –Law, is independent of film thickness. Typically in geophysics (e.g. lava flow) it is ca. 2:1 for lower viscosity minerals, [7] in rough agreement with our data. [1] This gives us a specific hypothesis to test in our films: If the Kittel Law is satisfied, the widths will increase by approximately x3 for a thickness increase of x10; whereas if the folding/buckling law is satisfied, it will increase linearly by x10.

It is not necessary to go as far afield as lava flows to examine analogous phenomena: Microscopic flow with folding or buckling is well studied in polymers [8,9] and for Au on Si. [10] These authors show that there is not just a threshold between folding and buckling, but that a longer sequence of events can occur with increasing strain: Wrinkles, folds, buckling, and finally de-lamination. Wrinkles are generally smooth and sinusoidal and energetically favored only where the substrate modulus is << that of the overlapping film, whereas folds are more abrupt and occur for film/substrate modulus ratios nearer unity.

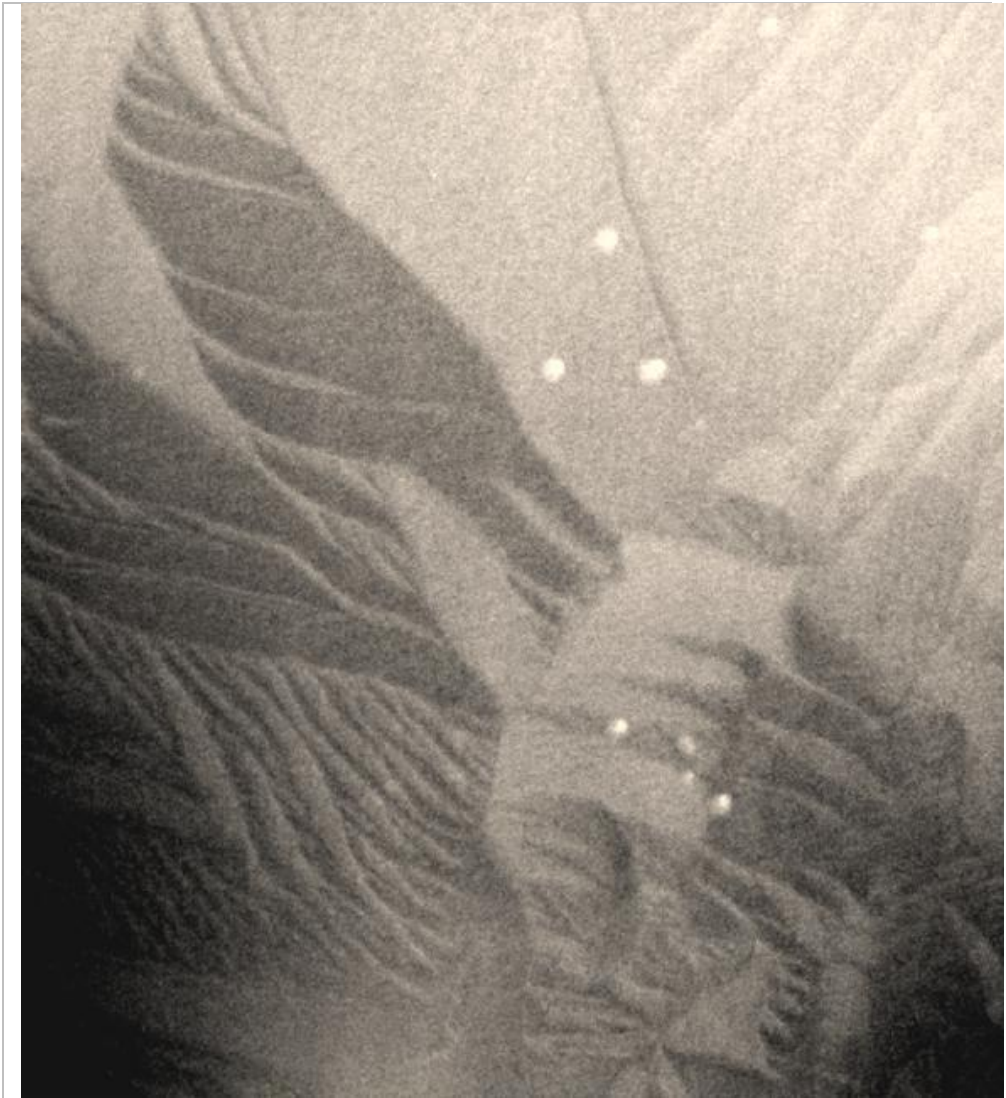

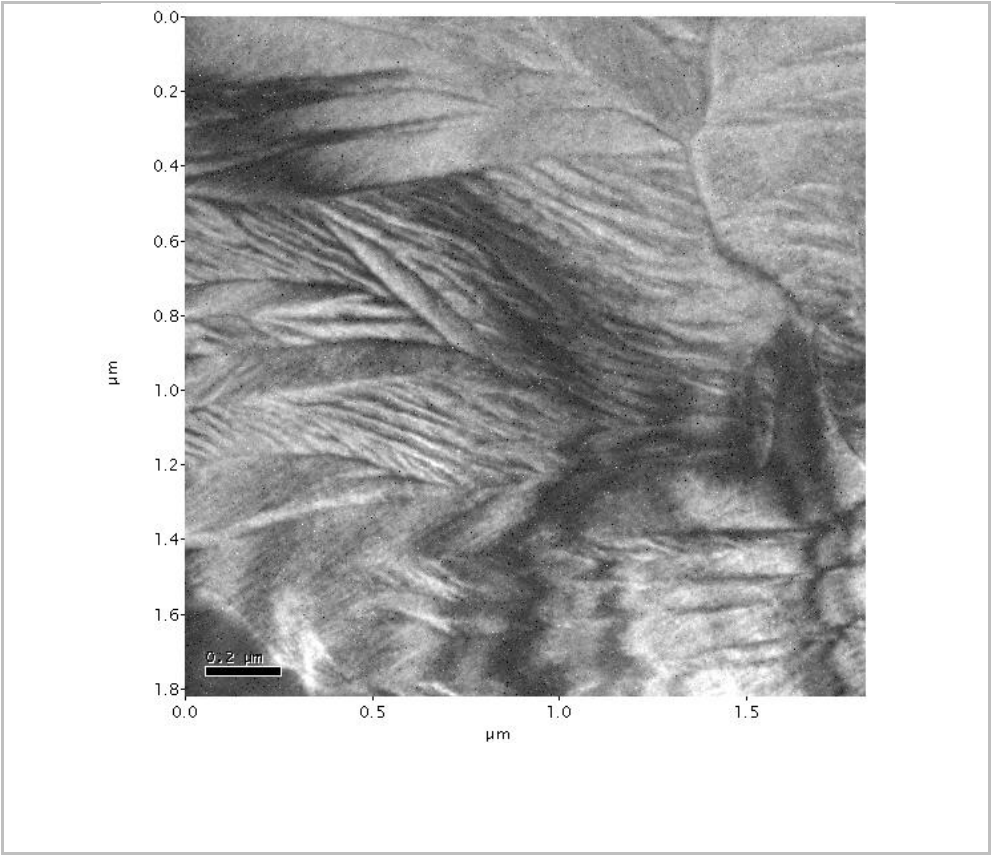

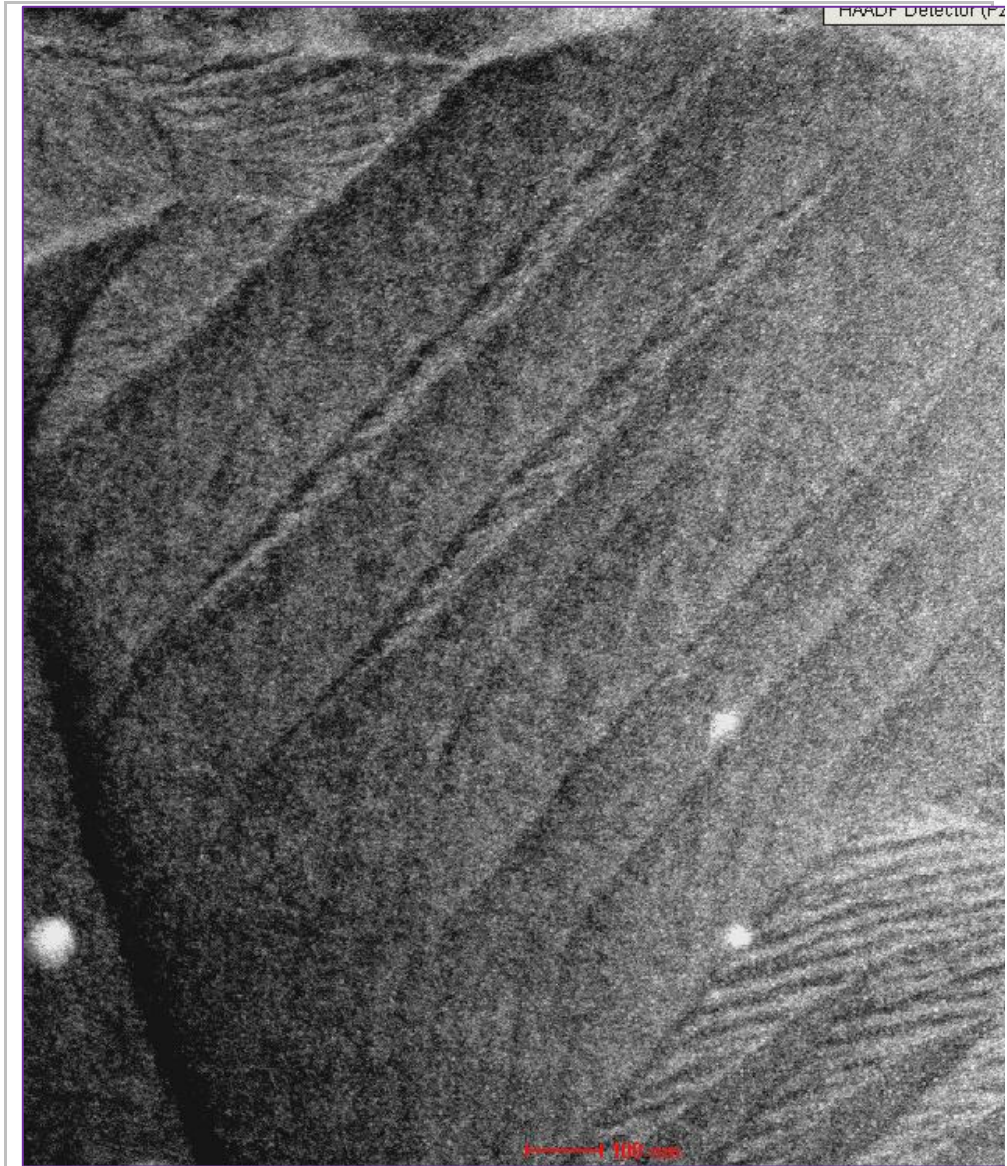

Figure 1 (a): TEM micrograph of $Pb(Fe_{1/2}Ta_{1/2})_xTi_yZr_{(1-x-y)}O_3$ showing folded domain structure analogous to ropy pahoehoe lava. Total image width: 2 μm. Bright spots are Ga from the focused ion beam (FIB). (b) Same as (a) but for a second specimen. (c) Here at higher resolution the 5-

10 nm wide parallel, straight ferroelectric stripes are clearly visible within the crescent-shaped ferroelastic domains.

-4-

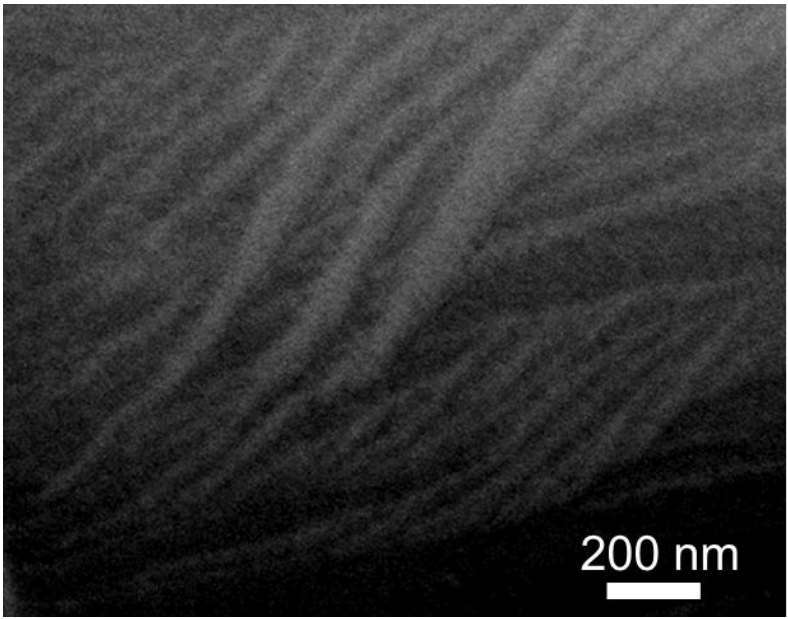

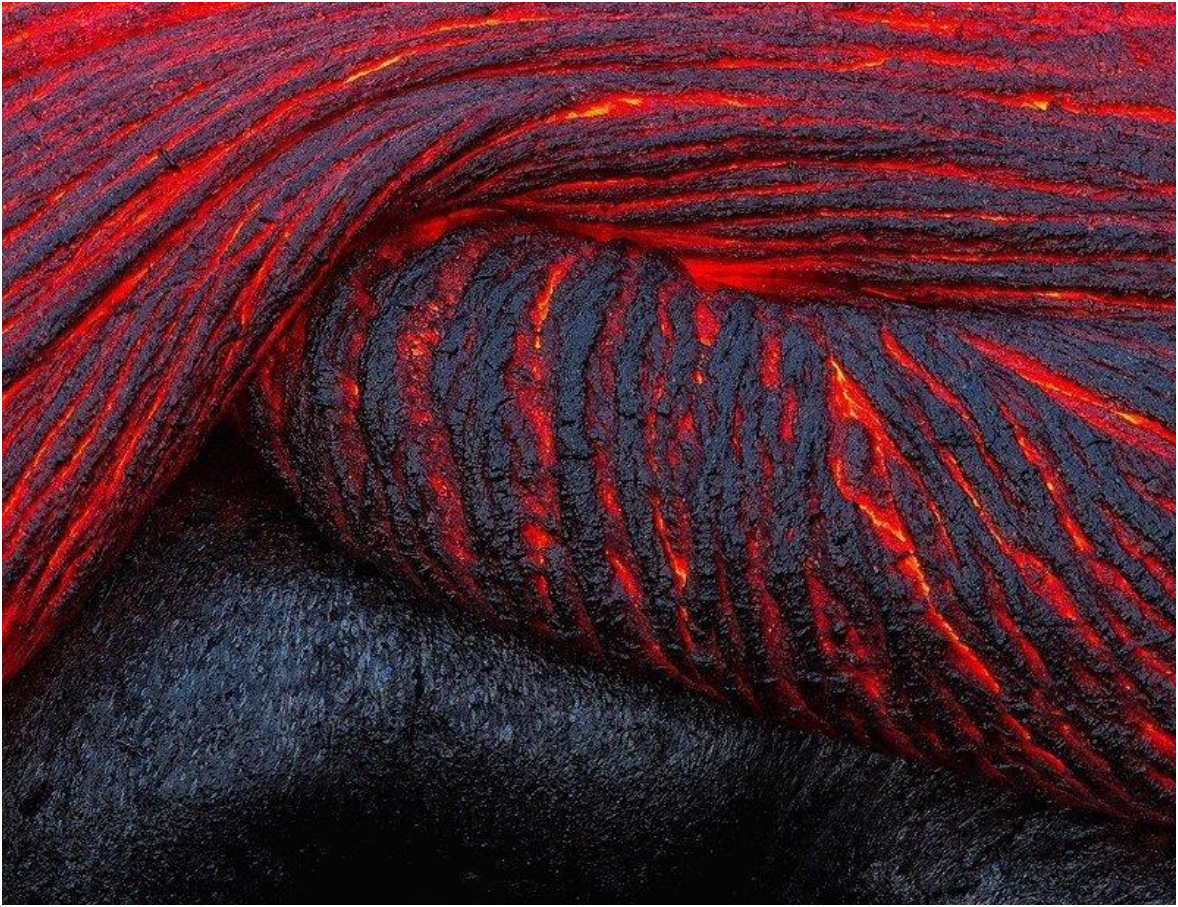

Figure 2 (a). (top) As in Figures 1 abc but for a third (thicker) specimen. Courtesy of Dr. Alina Schilling. (b) Macroscopic folds in Kilauea lava flow (scale is 1.0 m, a million times larger than (a)); courtesy of Justin Reznick (justinreznick.com; landscapephototours.com).

## 4. Folding models

Fold catastrophes are often described as saddle-node bifurcations. In one dimension the simplest case is given by

$$dx/dt = a - bx - cx^2 \qquad (1.)$$

in the limit of small b, and in two dimensions by that equation plus

$$dy/dt = -y. \qquad (2.)$$

Eq.(2) is just the Standard Linear Model for stress/strain, a combination of Maxwell and Kelvin-Voigt series and parallel springs and dashpots, which is known to apply to ferroelectric domain wall vertices under shear stress. [11]   That is, if y is strain, $\sigma$ is stress, $Y_1$ and $Y_2$, Young's moduli for two parallel responses in the equivalent springs and dashpot circuit, the Standard Linear Model is

$$dy/dt = [1/(Y_1+Y_2)](d\sigma/dt + Y_2\sigma/\eta - Y_2\, y)] = -ky \text{ for zero stress.} \qquad (3.)$$

This equation holds for domain interactions in various oxide ferroelectrics; a good example is proved in Fig. 3 for $BaTiO_3$ vertex-vertex collisions.

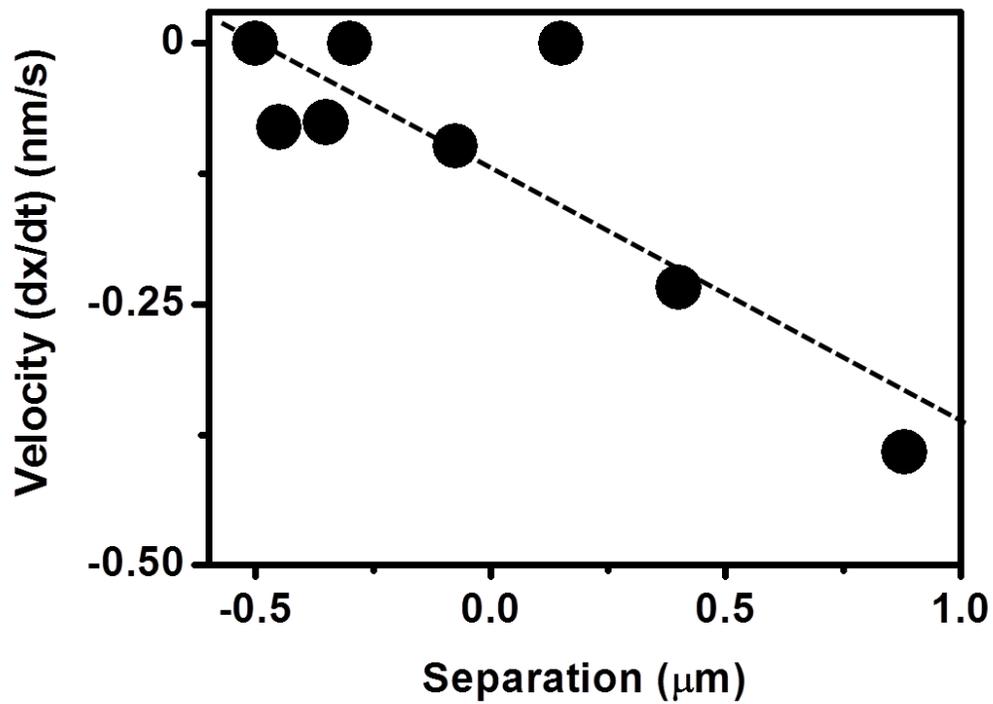

Fig. 3. Velocity V(x) versus x for domain vertices in barium titanate, confirming Eqs. 2 and 3.[11]

Eq.(1) is a nonlinear variant of Hooke's Law, with constant stress and a nonlinearity such as in the Ramberg-Osgood relationship; the latter approximates the power-law expansion of Eq.(1) by a single averaged exponent $<n>$, such that

$$dx/dt = a - bx - cx^{<n>},$$

(4.)

where for real materials, <n> is ca. 5.

So in the present context of domain walls and strain, this simple set of equations can reproduce saddle-point bifurcations and folding, provided $cx^2 \gg bx$. Schmalholtz and Schmid point out [12] that for simple shears, "Fold axial planes in the multi-layer are mostly curved and not parallel."

5. Effective viscosity

There is a possible connection between vortex creation in ferroelectric domains and the folding bifurcation discussed above: these vortices are very interesting new developments, [13-16] but when and why they occur is not fully understood. In this context we point out that vortex creation occurs in general only at high Reynolds numbers, whereas laminar flow (including folds) occurs at low Reynolds numbers. However, the Reynolds number for ferroelectric domain wall motion is unknown. Even defining a Reynolds number R requires the determination of a characteristic length L; this is usually done as:

$$R = vL/\eta. \qquad (5.)$$

However, in the present case this gives abnormally small numbers that are not easily related to viscosities in fluids.

High Reynolds numbers favor both vortex formation and strange attractors, and acute sensitivity to initial conditions. This suggests a line of study for vortex domains in ferroelectrics.

For ferroelectric thin films, the domain wall viscosity in Eq.(3) is given by

$$\eta = Fd/Av \qquad (6.)$$

where A is the domain wall area, and d is the domain thickness. It is not easy to estimate $\eta$ or R for ferroelectric domains. The science of nano-rheology of domain walls began with the work of Tybell, Paruch et al.[17,18] In our samples for a wall velocity v of ca. 1 nm/s [17,18] an area A of ca. $10^{-14}$ m$^2$ (100 nm per side), and d = 10 nm, if we assume a force per unit area F/A of 1.0 MPa from Massad and Smith's domain wall model [19] (higher than Salje's = 200 kPa for the low coercive force ferroelastic lead orthophosphate) [20] and a characteristic length L of 10-100 nm (domain width), we get a rough estimate of a wall viscosity of 2 x $10^6$ Poise ($10^8$ greater than water) and an extraordinarily small Reynolds number. The very small value of R and large viscosity $\eta$ remind us that Eq.3 describes smooth flow of a preexisting film over a substrate. But what is actually measured experimentally is usually a drag coefficient, not a Reynolds number. This suggests that it might be useful to redefine R from an equation other than Eq.(5). By incorporating a Prandl boundary layer, one finds that the dependence of drag coefficient (and hence F/A) upon Reynolds number can vary from linear to $R^{-1/5}$ as viscosity increases, which can imply orders of magnitude change. All of this makes it premature to try to define Reynolds numbers R for ferroelectric domain walls, or their characteristic lengths L, but suggests a new hydrodynamic direction for modeling domain wall dynamics.

Recent experiments on viscosity for nano-objects imply that viscosity is reduced by five or six orders of magnitude, compared with macroscopic amounts of the same substance.[22]

-7-

6. Implications for Room-temperature Multiferroic Memories

The domain-like features in Figures 1 and 2 switch polarization P with applied magnetic field H or electric field E at room temperature. [1] However, the magnetic switching fatigues after a few cycles, whereas switching with E does not fatigue. The present fold model suggests a qualitative explanation: It is known [21a] that the magnetoelectric coupling in this material is indirect via strain (magnetostriction plus piezoelectricity). However, the folds necessarily have very high local strains and will resist this striction-driven polarization reversal; this is a kind of fold-driven aging process. The implication is that although lead iron-tantalate zirconate-titanate is a good low-loss room-temperature multiferroic, commercialization of switching memory devices from it should minimize the domain folding patterns we see here.

7. Further theory:

Janovec and Privratskaya[23] have developed a theory of walls within walls. They point out that in ordinary optical microscopy of $BaTiO_3$ or other $ABO_3$ pseudo-cubic perovskites, such as that in the present paper, one recognizes ferroelastic domains with typical parallel planar walls. But that etching or piezoresponse force microscopy reveals that any ferroelastic domain consists of tiny non-ferroelastic ferroelectric domains. In the present context our rounded domain-like regions correspond to the ferroelastic walls in the Janovec-Privratska model, and the 5-10 nm ferroelectric domains are nested inside those (Fig.1c). To further summarize their $ABO_3$ description, which remains unpublished,[24] they maintain correctly that these phenomena are affected existence of an intermediate group 4/mmm between symmetry groups m3m and 4mm of the parent (prototype) and ferroic phase, respectively, m3m > 4/mmm > 4mm. Although the cubic phase m3m transforms directly to the ferroelectric tetragonal phase 4mm in a microscopic local sense, this phantom non-ferroelectric phase 4/mmm is "clearly revealed in the domain structure in form of ferroelastic domains" (our Figures 1 and 2). They write that these domains should be easily observable, whereas visualization of the non-ferroelastic ferroelectric domains will be difficult. However, high resolution PFM does show the interior ferroelectric nano-stripes, because the PFM technique gives black-and-white contrast to the ferroelectric nano-domains, whereas the TEM (albeit of high resolution) does not always permit such contrast, but Fig.1c does reveal these striped nano-domains. Thus we have realized the group theoretical prediction of Janovec and Privratska. We note however that the space group symmetry of the samples illustrated in Figs.1 and 2 here is R3c,[21a] although other specimens exhibited 4mm.[21b]   In general the ceramic specimens did not become R3m symmetry until below T = 250K (mm2 at ambient), whereas the free-standing FIB single nano-crystals were R3m at room temperature.   Such temperature shifts between ceramics and single crystal ferroelectrics are not uncommon and often arise from strain.

Other kinds of bifurcations are known in ferroelectrics in a different context:[25,26]  Temporal oscillations and period doubling are examples related to van der Pol, Lotka-Volterra, and other predator-prey models;[27] the former has been treated in terms of fold bifurcations.[28]  It is not known in the present work whether the ferroelastic folds arise from the PLD (pulsed laser deposition) deposition of the films or from the subsequent FIB (focused ion beam) thinning.  FIB is known to produce folds and wrinkles in polymer surfaces;[29]  and earlier work [30] shows a threshold electric field of 150 kV/cm in lead germanate for domain wall "wrinkling" instabilities.   In a different context very recent observations of vortex arrays of ferroelectric domains [15,16] strongly resemble the classic pictures[31] by Prandl of "von Karman vortex streets," which arise from ballistic objects passing through viscous media.

Summary

This work shows that the smooth round domains exhibited in TEM by FIB (focused ion-beam cut) single nano-crystals of lead iron-tantalate zirconate-titanate can be described as fold bifurcations, and suggests a shear strain term having velocity dx/dt

= a − cx$^2$ and dy/dt = -y; these are compatible with recent data. This explains the round domain edges and the failure of the Landau-Lifshitz-Kittel Law for domain widths. It also suggests the use of viscosity and an effective Reynolds number (yet to be defined) for such wall motion, and further hypothesizes that ferroelectric films can either have folds (low viscosity wall motion) or vortex structure [11-15] (high viscosity) but not both, which helps explain why vortex structures are not ubiquitous. Unpublished work by Janovec and Privratska strongly suggest that the micrographs in Figures 1 and 2 display nm-size rectilinear ferroelectric but non-ferroelastic domains nested inside larger rounded ferroelastic but non-ferroelectric domains; the smooth folds in the latter suggest saddle bifurcations due to domain wall viscosity, with a Ramsberg-Osgood model.[32] Commercial multiferroic memory devices may have improved fatigue behavior if ferroelastic folds can be minimized.

Acknowledgements: I thank Donald Evans and Marty Gregg for Figs. 1ab, Alina Schilling for Fig. 2a, Justin Reznick for Fig.2b, and Ray McQuaid for Fig.3.